\documentstyle[prl,aps,epsf,floats]{revtex}
\begin{document}
\draft
\twocolumn[\hsize\textwidth\columnwidth\hsize\csname @twocolumnfalse\endcsname
\title{Formation of a Heavy-Fermion State in 
       the 2D Periodic Anderson Model}
\author{Bumsoo Kyung}
\address{Max Planck Institute for Physics of Complex Systems, 
         Noethnitzer Str. 38, 01187 Dresden, Germany} 
\date{10 February 1998}
\maketitle
\begin{abstract}

   We study the formation of a heavy-fermion state in the 2D 
periodic Anerson model.
For $U=2$, the density of states, imaginary part of the self-energy
and effective magnetic moment all indicate the 
Kondo screening of local f electrons, 
leading to a coherent heavy-fermion state.
For $U=3$ and 4, the dominance of RKKY interaction over Kondo 
screening at low temperatures 
indicates a magnetic instability at zero temperature. 
A partial screening of magnetic moments, however, still gives rise to
a relatively sharp peak at the Fermi energy in the density of states.
\end{abstract}
\vskip2pc]
\narrowtext

   Since the discovery of the heavy-fermion materials
in rare-earth or actinide elements% 
\cite{Stewart:1984,Grewe:1991},
a possible dramatic transformation of 
localized $f$ electrons to 
a coherent heavy-fermion state has received considerable 
attention.
Experiments in these materials
have exhibited  
various unusual properties such as 
the huge coefficient in the magnetic susceptibility, specific 
heat and so on, which all indicate the formation of 
a coherent state with a large effective mass.
The periodic Anderson model\cite{Anderson:1961,Lee:1986}
has been considered as the most promising
candidate which might be able to describe
the anomalous features of these materials.  
The Hamiltonian for the 2D periodic Anderson model is 
\begin{eqnarray}
H & = & -t\sum_{\langle i,j \rangle \sigma}
        c^{+}_{i\sigma}c_{j\sigma} 
     +\sum_{i\sigma}[
        V(f^{+}_{i\sigma}c_{i\sigma} + 
          c^{+}_{i\sigma}f_{i\sigma})
       +\varepsilon_{f} f^{+}_{i\sigma}f_{i\sigma}] 
                                                         \nonumber  \\
 & + & U\sum_{i}
       (f^{+}_{i\uparrow}f_{i\uparrow}-n_{\uparrow})
       (f^{+}_{i\downarrow}f_{i\downarrow}-n_{\downarrow})  
                                                         \nonumber  \\
 & - & \mu \sum_{i\sigma}(c^{+}_{i\sigma}c_{i\sigma} 
                        +f^{+}_{i\sigma}f_{i\sigma}) 
                                               \; .
                                                           \label{eq1}
\end{eqnarray}
Here $t$ is the hopping parameter for $c$ electrons, $V$ 
the hybridization energy between $c$ and $f$ electrons, and 
$\varepsilon_{f}$ the energy level for $f$ electrons.
$U$ is the Coulomb repulsion energy for $f$ electrons,
$n_{\sigma}=\langle f^{+}_{i\sigma}f_{i\sigma} \rangle$
and $\mu$ is the chemical potential controlling the total electron
concentration.
$c_{i\sigma}$ and $f_{i\sigma}$ are annihilation operators for 
$c$ and $f$ electrons at site $i$ with spin $\sigma$, respectively.
When the hybridization term is eliminated by means of the 
equations of motion for $c$ and $f$ electrons,
the corresponding interacting Green's functions,
$G_{c}(\vec{k},\omega)$ and 
$G_{f}(\vec{k},\omega)$, 
are expressed in terms of the self-energy for $f$ electrons
$\Sigma_{f}(\vec{k},\omega)$
\begin{eqnarray}
G_{c}(\vec{k},\omega) & = & \frac{1}
                        {\omega-\varepsilon_{\vec{k}}+\mu
                         -\frac{V^{2}}
                          {\omega-\varepsilon_{f}+\mu
                           -\Sigma_{f}(\vec{k},\omega)}
                        }
                                               \; ,
                                                         \nonumber  \\
G_{f}(\vec{k},\omega) & = & \frac{1}
                        {\omega-\varepsilon_{f}+\mu
                         -\frac{V^{2}}
                          {\omega-\varepsilon_{\vec{k}}+\mu}
                         -\Sigma_{f}(\vec{k},\omega)
                        } 
                                               \; ,
                                                           \label{eq2}
\end{eqnarray}
where 
$\varepsilon_{\vec{k}} = -2t(\cos k_{x} + \cos k_{y})$ for nearest 
neighbor hopping.
As long as $f$ electrons are concerned, the interacting Green's 
function has the same structure as that for the usual Hubbard model with
the inverse of the noninteracting propagator 
$[G^{0}_{f}(\vec{k},\omega)]^{-1}$
replaced by
$\omega-\varepsilon_{f}+\mu
  -\frac{V^{2}}
  {\omega-\varepsilon_{\vec{k}}+\mu}$.
For later use, we also show the noninteracting ($U=0$) energy dispersion 
due to hybridization,
\begin{equation}
E^{0}_{\pm}(\vec{k})  =  \frac{(\varepsilon_{f}+\varepsilon_{\vec{k}}
                         -2\mu) \pm
                         [(\varepsilon_{f}-\varepsilon_{\vec{k}})^{2}
                          +4V^{2}]^{1/2}}{2}
                                               \; .
                                                           \label{eq2-1}
\end{equation}

   Recently the infinite\cite{Jarrell:1993} and  
two\cite{Vekic:1995} dimensional periodic Anderson 
models for the symmetric case 
have been studied by using quantum Monte Carlo (QMC) calculations.
In these works for a half-filled band,
the authors found a strong competition between 
the Kondo effect\cite{Kondo:1964,Hewson:1993}  
and antiferromagnetic order depending on the 
size of $V$ and $U$.
Because of its symmetric nature, however, the model can not describe 
any possible formation of 
a coherent heavy-fermion state from localized
$f$ electrons.
The 2D asymmetric periodic Anderson model has been investigated 
by McQueen {\it et al.}\cite{McQueen:1993}
within fluctuation exchange (FLEX) approximation. 
These authors found a developing density of states near the Fermi 
energy with decreasing temperature, which was argued as an 
indication of the Kondo resonance.
However, they could not observe a Fermi liquid behavior in the 
calculations of the self-energy.
This is possibly due to the insufficient treatment of strong 
local correlations and 2D spin fluctuations
in the FLEX approximation%
\cite{Vilk:1997}.

   Recently we have developed a new approximation scheme for the 2D 
Hubbard model which shows 
the essential features of the model%
\cite{Kyung:19971}.
This theory satisfies
simultaneously
the correct atomic limit for large frequencies (reflected as 
the correct asymptotic behavior
of the self-energy at large $\omega$) as well as  
2D spin fluctuations due to
the Mermin-Wagner theorem%
\cite{Mermin:1966}. 
These two features are very important in the study of the 2D  
periodic Anderson model, 
in that the screening 
of local moments by conduction electrons is properly described by 
the first property and 
any possible magnetic instability at zero temperature
by the second.
One more good feature of our formulation 
over FLEX approximation or second-order (in $U$)
perturbation study is that 
the dynamical spin susceptibility can be computed accurately 
by imposing the exact sumrules. 
In this Letter we present the strong numerical evidence for 
the formation of a coherent heavy-fermion 
state from localized $f$ electrons by showing the density of states,
low frequency behavior of the self-energy, effective magnetic 
moment and quasiparticle residue.

   In order to compute properly 
the spin, charge, and particle-particle susceptibilities which 
govern the interactions between electrons,
we impose the following three exact sumrules to 
them%
\cite{Kyung:19971,Vilk:1997}:
\begin{eqnarray}
\frac{T}{N}\sum_{q}\chi_{sp}(q) & = & n-2\langle n_{\uparrow}n_{\downarrow}
                                         \rangle
                                                         \nonumber  \\
\frac{T}{N}\sum_{q}\chi_{ch}(q) & = & n+2\langle n_{\uparrow}n_{\downarrow}
                                         \rangle-n^{2}
                                                         \nonumber  \\
\frac{T}{N}\sum_{q}\chi_{pp}(q) & = & \langle n_{\uparrow}n_{\downarrow}
                                         \rangle
                                               \; .
                                                           \label{eq3}
\end{eqnarray}
$T$ and $N$ are the absolute temperature and 
number of lattice sites. 
$q$ is a compact notation for $(\vec{q},i\nu_{n})$ where
$i\nu_{n}$ are either Fermionic or 
Bosonic Matsubara frequencies. 
The dynamical spin, charge and particle-particle susceptibilities are 
calculated by 
\begin{eqnarray}
\chi_{sp}(q)&=&\frac{2\chi^{0}_{ph}(q)}{1-U_{sp}\chi^{0}_{ph}(q)}
                                                         \nonumber  \\
\chi_{ch}(q)&=&\frac{2\chi^{0}_{ph}(q)}{1+U_{ch}\chi^{0}_{ph}(q)}
                                                         \nonumber  \\
\chi_{pp}(q)&=&\frac{ \chi^{0}_{pp}(q)}{1+U_{pp}\chi^{0}_{pp}(q)}
                                               \; .
                                                           \label{eq4}
\end{eqnarray}
$\chi^{0}_{ph}(q)$ and 
$\chi^{0}_{pp}(q)$ are 
irreducible particle-hole and particle-particle susceptibilities, 
respectively, which are computed from
\begin{eqnarray}
\chi^{0}_{ph}(q) & = & - \frac{T}{N}\sum_{k}G_{f}^{0}(k-q)G_{f}^{0}(k)
                                                         \nonumber  \\
\chi^{0}_{pp}(q) & = &  \frac{T}{N}\sum_{k}G_{f}^{0}(q-k)G_{f}^{0}(k)
                                               \; ,
                                                           \label{eq5}
\end{eqnarray}
where 
$G_{f}^{0}(k)$ is the noninteracting Green's function for 
$f$ electrons defined earlier.
$U_{sp}$, $U_{ch}$, and $U_{pp}$ in Eq.~\ref{eq4} are    
renormalized interaction constants for each channel which are 
calculated self-consistently 
by making an ansatz
$U_{sp} \equiv U\langle n_{\uparrow}n_{\downarrow} \rangle/
(\langle n_{\uparrow} \rangle
\langle n_{\downarrow} \rangle)$\cite{Vilk:1997}
in Eq.~\ref{eq3}.
By defining 
$U_{sp}$, $U_{ch}$, and $U_{pp}$     
this way, the Mermin-Wagner 
theorem as well as the correct
atomic limit for large $\omega$ are satisfied 
simultaneously\cite{Kyung:19971}. 
%
%
%In order to find the chemical potential for interacting electrons,
%first
%we calculate
%Eqs.~(\ref{eq3})-(\ref{eq5}) and the self-energy 
%(Eqs.~(\ref{eq1}) in Ref.~\onlinecite{Kyung:19971})
%with the noninteracting Green's function whose 
%noninteracting chemical potential gives 
%a desired electron concentration. 
%Then, 
%the chemical potential for interacting electrons is determined 
%in such a way that the calculated electron concentration 
%with the interacting Green's function 
%is the same as the desired value.
%
%
Throughout the calculations we fixed 
$\varepsilon_{f}=0.45$, $V=1$ and 
the total electron concentration 
at 2.25.
The unit of energy is $t$ and all energies 
are measured from the chemical potential $\mu$. 
We used a $64 \times 64$ lattice in momentum space 
and performed the calculations by means of 
well-established fast Fourier transforms
(FFT).
It should be also noted that we used a real frequency formulation
in Eqs.~(\ref{eq2})-(\ref{eq5}) to avoid any possible uncertainties
associated with numerical analytical continuation.

\begin{figure}
 \vbox to 7.5cm {\vss\hbox to -5.0cm
 {\hss\
       {\includegraphics{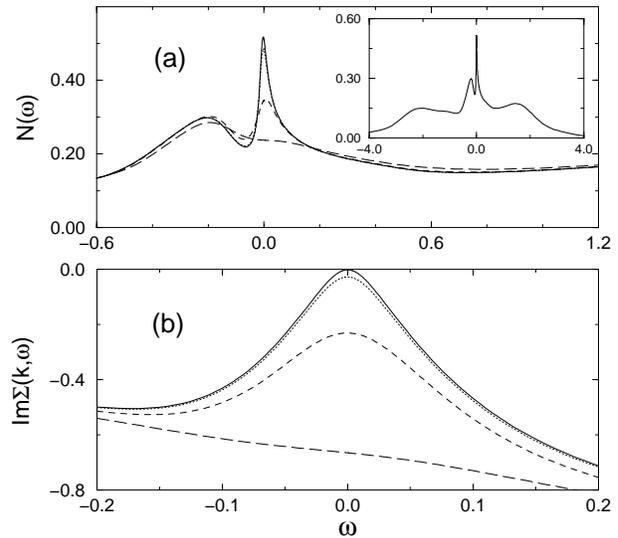}
       }
  \hss}
 }
\caption{(a) Density of states and 
         (b) imaginary part of the self-energy at the Fermi surface
             for $U=2$, $\varepsilon_{f}=0.45$ and $V=1$ with
             decreasing temperature.
             The long-dashed, dashed, dotted and solid curves
             denote
             $T=1/16$, 1/64, 1/256 and 1/1024, respectively.
             The inset in (a) shows the density of states
             for $T=1/1024$
             in the wide range of frequency axis.}
\label{fig1}
\end{figure}

   We start in Fig.~\ref{fig1} by studying 
the density of states 
(Fig.~\ref{fig1}(a)) and  
imaginary part of the self-energy
(Fig.~\ref{fig1}(b)) for $U=2$ with decreasing temperature
from $T=1/16$ to 1/1024.
As the temperature is decreased, a sharp peak develops at the 
Fermi energy in Fig.~\ref{fig1}(a).
At $T=1/16$ this peak is completely absent.
To find that this feature is associated with
the onset of the Fermi liquid, we calculated the imaginary part of the 
self-energy  
in Fig.~\ref{fig1}(b). 
With decreasing temperature, the scattering rates progressively 
decrease and vanish at the Fermi energy much faster than linearly
in frequency.
The log-log plot of the scattering rates vs. frequency
for $T=1/1024$ shows 
$Im \Sigma(\vec{k}_{F},\omega) \sim \omega^{1.62}$ 
in a relatively narrow interval
of $\pm [0.005-0.025]$. 

\begin{figure}
 \vbox to 7.5cm {\vss\hbox to -5.0cm
 {\hss\
       {\includegraphics{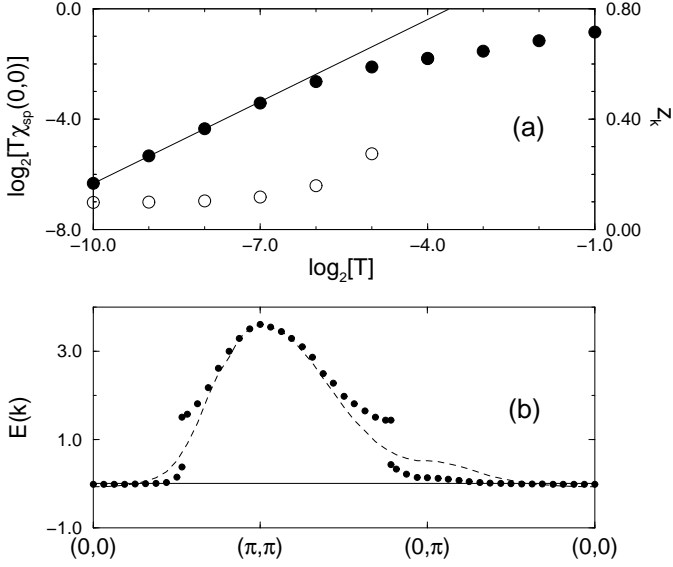}
       }
  \hss}
 }
\caption{(a) Effective magnetic moment (filled circles) and  
             quasiparticle residue (open circles), and 
         (b) quasiparticle dispersion (small filled circles) for 
             $T=1/256$
             against noninteracting dispersion (dashed curve) 
             in the upper branch 
             for $U=2$, $\varepsilon_{f}=0.45$ and $V=1$.
             The solid line in (a) is a linear interpolation 
             at low temperatures.}
\label{fig2}
\end{figure}

   In order to establish a more firm ground that this is indeed
due to the screening 
of magnetic moments by conduction electrons,  
we show the effective magnetic moment defined as 
$T\chi_{sp}(0,0)$ upon decreasing the temperature 
(filled circles in Fig.~\ref{fig2}(a)).
Below $T=1/128$, the effective magnetic moment vanishes linearly 
in temperature, a clear indication of 
the Kondo screening.
Above this temperature, it
deviates significantly from a straight line 
and appears gradually saturating at high temperatures.
The quasiparticle residue z$_{\vec{k}}$ at the Fermi surface
along the direction $(0,0)-(\pi,\pi)$
is also presented as the open circles
in Fig.~\ref{fig2}(a).
Below this same temperature, it saturates at 0.1.
Hence, the effective mass of the 
quasiparticle becomes ten times heavier than the bare electron mass
at low enough temperatures.  
The quasiparticle residue is almost isotropic throughout the 
Fermi surface.
In Fig.~\ref{fig2}(b) we present the drastic change of the 
quasiparticle dispersion (small filled circles)
for $T=1/256$ near the Fermi energy.
The dashed curve is the noninteracting dispersion in the upper 
branch, 
$E^{0}_{+}(\vec{k})$ in Eq.~\ref{eq2-1},
where the Fermi energy stays.
The interacting quasiparticle dispersion becomes dramatically 
flattened along the directions $(0,0)-(\pi,\pi)$ and 
$(0,\pi)-(0,0)$, leading to a heavy effective mass as well as
large density of states at the Fermi energy.
It is of interest to notice that the interacting energy band appears
separated into f electron (lower band) and c electron
(upper band) dominating parts, which is absent in the 
noninteracting band.

\begin{figure}
 \vbox to 6.5cm {\vss\hbox to -5.0cm
 {\hss\
       {\includegraphics{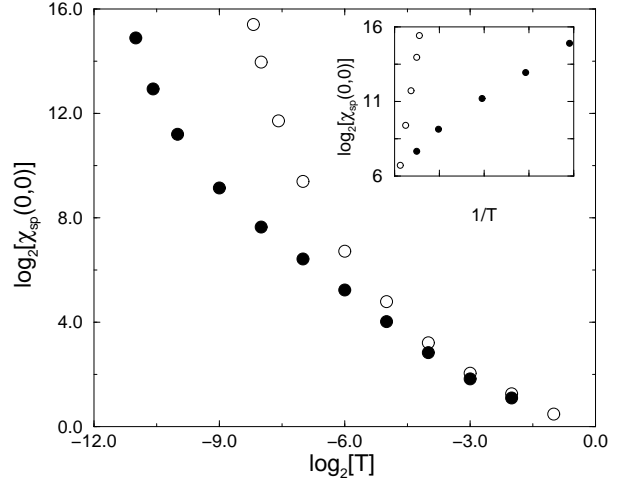}
       }
  \hss}
 }
\caption{ Static spin susceptibilities at $\vec{q}=(0,0)$ 
          for $U=3$ (filled circles) and 
          $U=4$ (open circles).  
          The inset shows the temperature dependence of 
          $\chi_{sp}(0,0)$ in the low temperature regime.
          Notice the logarithmic scale for $\chi_{sp}(0,0)$.}
\label{fig3}
\end{figure}

   In order to understand the role of $U$ in the one and two particle 
properties,
we also performed 
the calculations for $U=3$ and 4 with all the 
other parameters unchanged. 
In both cases, we found the indication of a ferromagnetic instability with 
decreasing temperature, which is signaled by a divergent 
behavior of 
$\chi_{sp}(\vec{q},0)$ at $\vec{q}=(0,0)$ 
in Fig.~\ref{fig3}.
Note that there is no finite temperature phase transition 
in two dimensions
due to the Mermin-Wagner theorem.
The inset clearly shows
$\chi_{sp}(0,0) \sim \exp(constant/T)$ at low temperatures.
This magnetic instability 
happens because the RKKY interaction between f electrons becomes
dominating over the Kondo screening for large $U$%
\cite{Tsunetsugu:1993}.

\begin{figure}
 \vbox to 7.5cm {\vss\hbox to -5.0cm
 {\hss\
       {\includegraphics{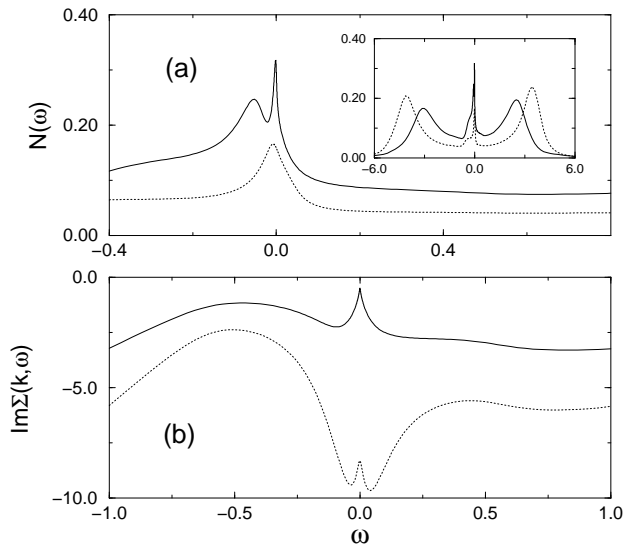}
       }
  \hss}
 }
\caption{(a) Density of states and 
         (b) imaginary part of the self-energy 
             at the noninteracting Fermi surface
             for $\varepsilon_{f}=0.45$ and $V=1$ 
             with $U=3$, $T=1/2048$ (solid curves) and 
             $U=4$, $T=1/290$ (dotted curves). 
             The inset in (a) shows the density of states
             in the wide range of frequency axis.}
\label{fig4}
\end{figure}

   In Fig.~\ref{fig4} 
we present the density of states 
(Fig.~\ref{fig4}(a)) and  
imaginary part of the self-energy
(Fig.~\ref{fig4}(b)) for $U=3$, $T=1/2048$ (solid curves) and 
$U=4$, $T=1/290$ (dotted curves).  
These temperatures are the lowest ones for each $U$ where  
we found the convergent solutions for $U_{sp}$ in our 
numerical calculations.
The inset in (a) shows
the density of states in the wide range of frequency axis.
Due to 2D strong spin fluctuations at low temperatures,
the spectral weight is significantly suppressed in the  
low frequency regime
(inset in Fig.~\ref{fig4}(a)), compared with that for $U=2$
(inset in Fig.~\ref{fig1}(a)). 
To our surprise,
a relatively sharp peak is still persisting at the Fermi energy
even in the presence of strong spin fluctuations.
In order to understand the nature of this peak, 
we computed the scattering rates
in Fig.~\ref{fig4}(b).
$Im \Sigma(\vec{k}_{F},\omega)$ shows a small dip at the Fermi energy in 
a large scattering background, which is 
different from the case for $U=2$.
This local minimum of the scattering rates 
at the Fermi energy is responsible for the 
sharp structure in the density of states.
This feature comes from a {\it partial} screening of f electrons by 
conduction electrons for the following reason.
In the critical region the imaginary part of the spin susceptibility 
becomes singular like
$Im \chi_{sp}(\vec{q},\nu) \sim \delta(\nu-\Omega)$ 
near $\vec{q}=(0,0)$
for a ferromagnetic case.
$\Omega$ is the characteristic spin fluctuation energy which is 
less than $10^{-4}$ for the parameters used for $U=3$ and 4.
Hence, the imaginary part of the self-energy near the Fermi energy 
is given by 
\begin{eqnarray}
Im \Sigma_{f}(\vec{k}_{F},\omega) & \sim & \sum_{\vec{q}}
                        \int d\nu \;
                        Im \chi_{sp}(\vec{q},\nu)  
                        [B(\nu)+F(\nu-\omega)]
                                                         \nonumber  \\
                & \times & Im G^{0}_{f}(\vec{k}_{F}-\vec{q},\omega-\nu)  
                                                         \nonumber  \\
                & \sim & 
                     \sum_{\vec{q}}
                     \frac{1}{2}
           (1+\frac{\varepsilon_{f}-\varepsilon_{\vec{k}_{F}-\vec{q}}}
           {\sqrt{(\varepsilon_{f}-\varepsilon_{\vec{k}_{F}-\vec{q}})^{2}
                         +4V^{2}}})
                                                         \nonumber  \\
                & \times &   
                     \delta(\omega-E^{0}_{+}(\vec{k}_{F}-\vec{q}))
                                               \; .
                                                           \label{eq6}
\end{eqnarray}
Since 
$Im \chi_{sp}(\vec{q},\nu)$ shows a nearly divergent behavior
in the small region with the radius of approximately 
$\pi/16$ from the origin in our calculations,
the summation of $\vec{q}$ is over this area.
Due to the mixture of f and c electrons through $V$, 
generally the hybridized band 
$E^{0}_{+}(\vec{k})$ has some dispersion or slope 
at the noninteracting Fermi surface.
Because of this feature, the contributions 
from $\vec{q} \neq (0,0)$ give rise to larger scattering rates
at $\omega \neq 0$ than those 
at $\omega = 0$, leading to a small dip at the Fermi energy
in the scattering rates.  
This is in turn responsible for a relatively 
sharp peak in the density of states.
Hence,
a peak at $\omega=0$ in the density of states for $U=3$ and 4
is caused by {\it mixing} of f and c electrons, that is, a {\it partial} 
screening of local moments by conduction electrons.
To confirm this argument, we also calculated the density of states
and imaginary part of the self-energy by increasing the 
hybridization strength $V$ with all the other parameters including 
the total electron concentration fixed.  
\begin{figure}
 \vbox to 7.5cm {\vss\hbox to -5.0cm
 {\hss\
       {\includegraphics{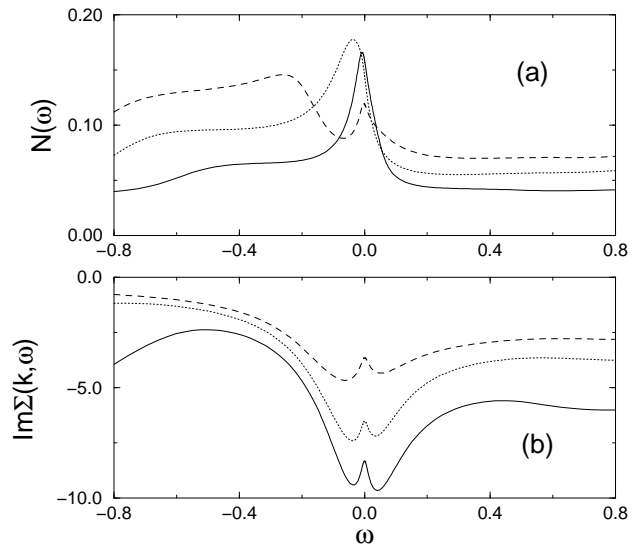}
       }
  \hss}
 }
\caption{(a) Density of states and 
         (b) imaginary part of the self-energy
             at the noninteracting Fermi surface
             for $\varepsilon_{f}=0.45$, $U=4$ and 
             $T=1/290$ with increasing $V$.
             $V=1$, 1.25 and 1.5 are denoted as the solid, 
             dotted and dashed curves, respectively.}
\label{fig5}
\end{figure}
As mixing ($V$) is increased, the scattering rates from  
spin fluctuations are more suppressed near the Fermi energy
in Fig.~\ref{fig5}(b).
The distance between the local maximum and the dip is given 
by the value
$E^{0}_{+}(\vec{k}_{F}-\vec{q})$ at 
$q=\pi/16$ for all three different $V$.
In a hypothetical situation where 
$E^{0}_{+}(\vec{k})=constant$ (no mixing) and 
the same $Im \chi_{sp}(\vec{q},\nu)$ is taken,
the scattering rates would be maximum at the Fermi energy,
leading to a {\it pseudogap} instead of a {\it peak} in the density of 
states.
%
%
%
%   Before closing we comment the dependence of the magnetic 
%ground state
%on the total electron concentration. We increased $n_{T}$ 
%up to 3.30 for $U=4$.
%Up to 2.50 the ferromagnetic ground state persists and beyond
%this value the ground state becomes paramagnetic until another 
%magnetic (antiferromagnetic) ground state is established
%around $n_{T}=3$. Beyond this concentration again the paramagnetic ground 
%state reappears.
%Near $n_{T}=3$ the Fermi surface in the upper branch of dispersion
%$E^{0}_{+}(\vec{k})$ is very similar to that for a half-filled 
%2D band.
%
%

   In summary, 
the formation of a coherent heavy-fermion state in the 2D 
periodic Anerson model has been studied on the basis of the recently 
developed theory for the Hubbard model.
For $U=2$, the growing density of states and 
rapidly vanishing scattering rates near the Fermi energy as well as 
linearly vanishing (in temperature) 
effective magnetic moment below a characteristic temperature,
strongly support
the Kondo screening 
of f electrons,  
leading to a heavy-fermion state.
For $U=3$ and 4, the dominance of RKKY interaction over Kondo 
screening at low temperatures 
indicates a magnetic instability at zero temperature. 
The persistence of a peak 
at $\omega=0$ 
even in the presence of strong spin fluctuations
is due to a partial
screening of f electrons by conduction electrons.

    The author would like to thank Prof. P. Fulde,  
and Drs. S. Blawid, R. Bulla, G. Kaliullin,
M. Laad, W. Stephan, P. Thalmeier and numerous other colleagues 
in the Max Planck Institute
for Physics of Complex Systems 
for useful discussions.   
\end{document}